\documentclass{appolb}
\usepackage{epsfig}
\usepackage{psfig}
\newcommand{\cp}{\mbox{$\not \hspace{-0.15cm} C\!\!P \hspace{0.1cm}$}}
\def\greaterthansquiggle{\raise.3ex\hbox{$>$\kern-.75em\lower1ex\hbox{$\sim$}}}
\def\lessthansquiggle{\raise.3ex\hbox{$<$\kern-.75em\lower1ex\hbox{$\sim$}}}
\def\gl{\raise.3ex\hbox{$<$\kern-.68em\lower1ex\hbox{$>$}}}

\newcommand{\lts}{\lessthansquiggle}
\newcommand{\imag}{\Im {\rm m}}
\newcommand{\real}{\Re {\rm e}}
\begin{document}
\eqsec  % uncomment this line to get equations numbered by (sec.num)
\title{Supersymmetry at the PLC
\thanks{Presented at PLC2005, Kazimierz, September 5-8,2005.}%
}
\author{Rohini M. Godbole
\address{Center for High Energy Physics, Indian Institute of Science, Bangalore,560012, India.}
}
\maketitle
\begin{abstract}
In this talk I will begin with a very brief discussion as to why TeV scale
Supersymmetry forms an important subject of the studies at all the current and
future Colliders. Then, I will give different examples where the Photon Linear 
Collider, PLC, will be able to make unique contributions.  PlC's most
important role is in the  context of Higgs Physics, due to its
ability of  accurate determination of $\Gamma_{\gamma \gamma}$
as well as the possibilities it offers for the determination of the CP property
of the Higgs boson and of possible  CP mixing in the Higgs sector. Further, 
the PLC can provide probes of SUSY in the regions of the SUSY parameter space, 
which are either difficult or inaccessible at the LHC and also in the $e^+e^-$ 
mode of the International Linear Collider (ILC). 

\end{abstract}
\PACS{11.30.Pb,12.60.Jv,14.80.Cp}
  
\section{Introduction}
In this talk I want to discuss  the special role that the Photon 
Linear Collider (PLC) can play when it comes to Supersymmetry searches/studies 
at the future colliders. Before doing this, let us just briefly recapitulate
the basics of  Supersymmetry (SUSY), the attractions that  the TeV scale
supersymmetry holds for the Particle Physics community and the reasons 
why  the searches 
for SUSY form a significant part of the physics studies at the colliders:
currently running and/or in planning/construction\cite{SUSYbook}. 
Supersymmetry, a symmetry transformation between fermions and bosons, is the
only possible extension of the space-time symmetries to particle interactions. 
In other words this is the only consistent way to combine space-time symmetries
with an internal symmetry. In addition Supersymmetric field theories are the 
only quantum field theories which remain 'natural'\cite{kaul} even in presence
of scalars. As a result Supersymmetry helps stabilize the EW symmetry breaking 
scale against radiative corrections.  SUSY thus provides a  solution to 
the 'naturalness' problem, which is theoretically very attractive and elegant.
In these theories,  associated with every standard model particle there is a
supersymmetric partner, the sparticle,  differing in spin by 1/2.  The left 
panel in 
\begin{figure}[htb]
\includegraphics*[scale=0.5]{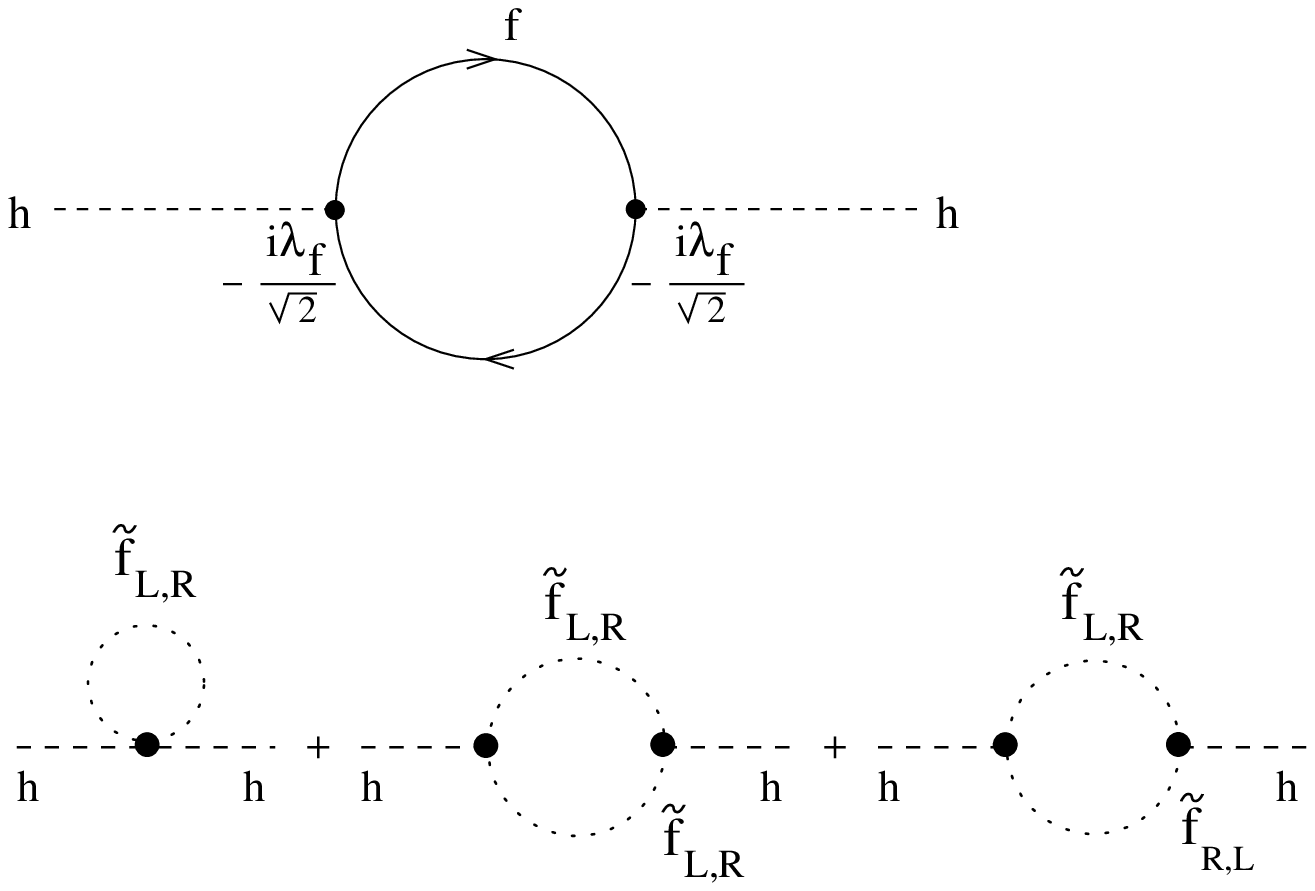}
\includegraphics*[scale=0.30]{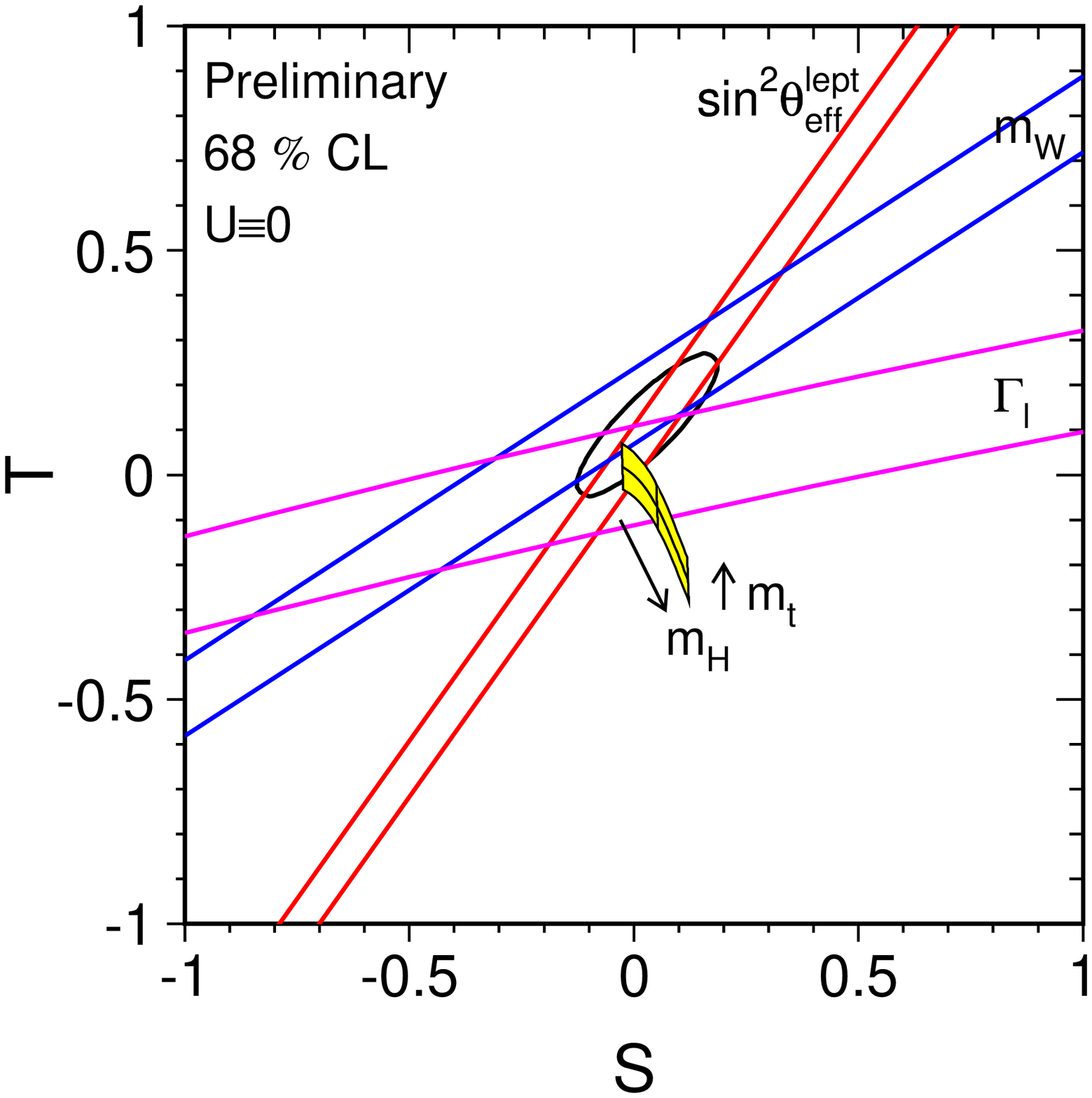}
\caption{Stabilisation of Higgs mass against radiative corrections and
experimental evidence for a weakly coupled light Higgs.\label{fig1}}
\end{figure}
Fig.~\ref{fig1} indicates how the sparticle loops help cancel the large
self energy corrections, keeping the Higgs mass 'naturally' light. As 
a matter of fact in the limit of perfect supersymmetry, where the particle 
and sparticle masses are equal, these corrections will cancel each other 
exactly. Even if SUSY is broken, one can show that existence of a TeV scale 
supersymmetry will keep the Higgs naturally light.

The experiments of the past few decades, culminating in the high precision 
measurements at the colliders and the neutrino experiments, have established
the correctness of both the gauge sector and the flavour sector of the SM 
Lagrangian given by,
\begin{equation}
\mathcal{L}=-{\frac{1}{4}  F^a_{\mu\nu} F^{a\, \mu\nu}
+ i \bar \psi  \not\!D \psi}  
+ {\psi^T \lambda \psi h + h.c. } 
+ { |D_{\mu} h|^2 - V(h)}.
\end{equation}
Only the scalar sector remains without direct evidence.  The Tevatron and the 
LEP/SLC give 'indirect' bounds on the Higgs mass. Analysis of precision 
measurements from LEP in terms of the Oblique parameters, $S,T,U$\cite{peskin},
constrain strongly any {\bf nondecoupling NEW} physics beyond the SM. The 
plot in the right panel of Fig.~\ref{fig1}, taken from the 
http://lepewwwg.web.cern.ch, illustrates these constraints. 
This indirect upper bound on the Higgs mass at $95\%$ c.l. is $251$ GeV, 
whereas the direct searches give a lower 
bound of $114$ GeV. Thus the precision measurements like a 'light' Higgs. As
a matter of fact, theorists like a 'light' Higgs as well. If the SM is an 
effective theory, then we expect $180 < m_h < 200$ GeV . Further, in a 
model independent analysis~\cite{barbieri}, one can show that if the scale
for New Physics $\Lambda_{NP} < 10$ TeV, then one expects, demanding
'naturalness' $195 < m_h < 215$; SUSY being a particular example of the
New Physics which keeps the Higgs 'naturally' light. These experimental 
indications of a 'light' Higgs make SUSY theoretically  very attractive. 
The search for SUSY is thus the case of experiments chasing a beautiful
theoretical idea. Even if it is a symmetry of nature, it is clearly broken. 
Further, all the experimental searches so far have yielded only  negative
results, giving only {\bf lower} limits on the sparticle masses. The only,
{\it very indirect} indication for SUSY at present seems to be the absence 
of the unification of the three gauge couplings in the SM, whereas in the
MSSM the three do unify. 
\begin{figure}
\includegraphics*[scale=0.38]{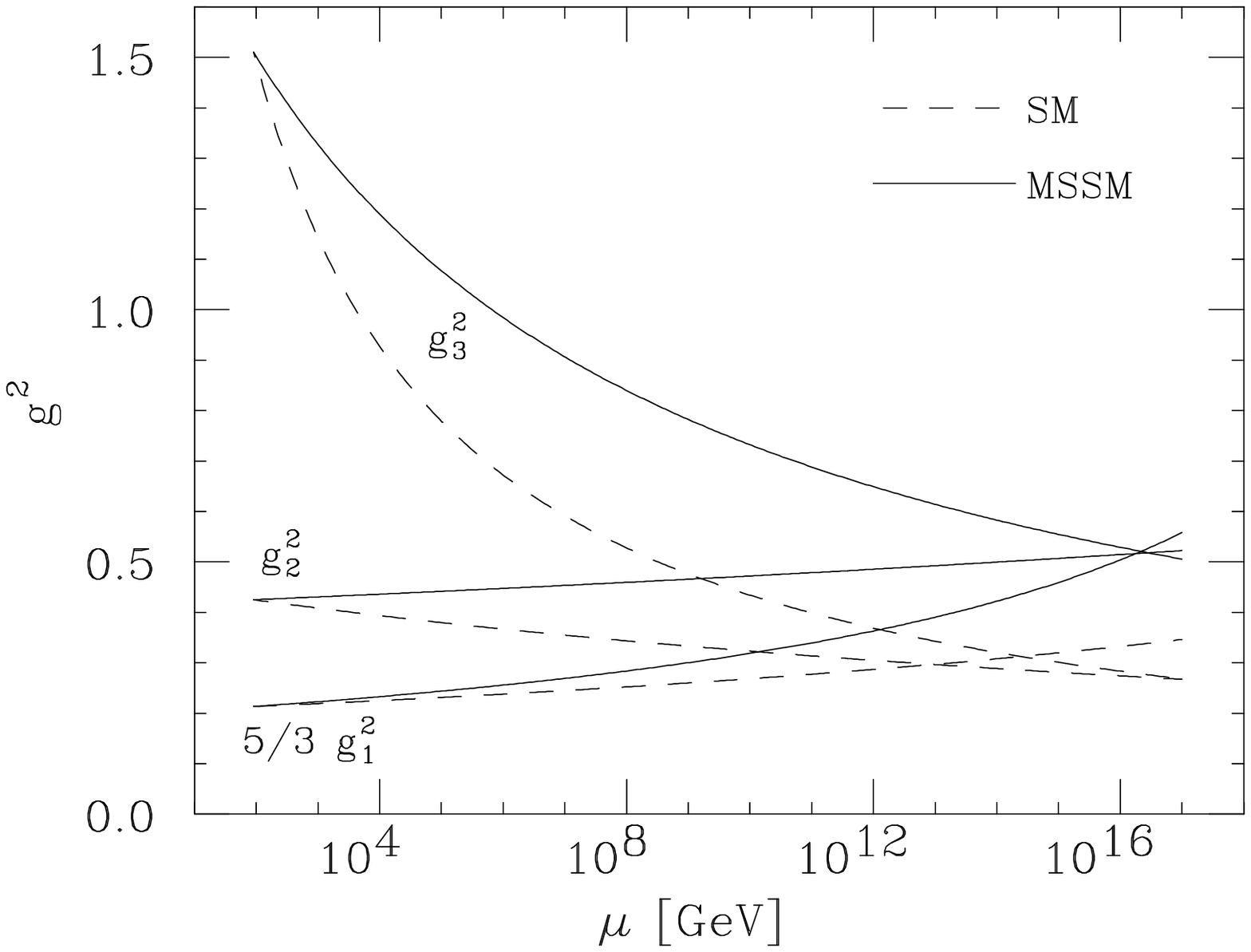}
\includegraphics*[scale=0.30] {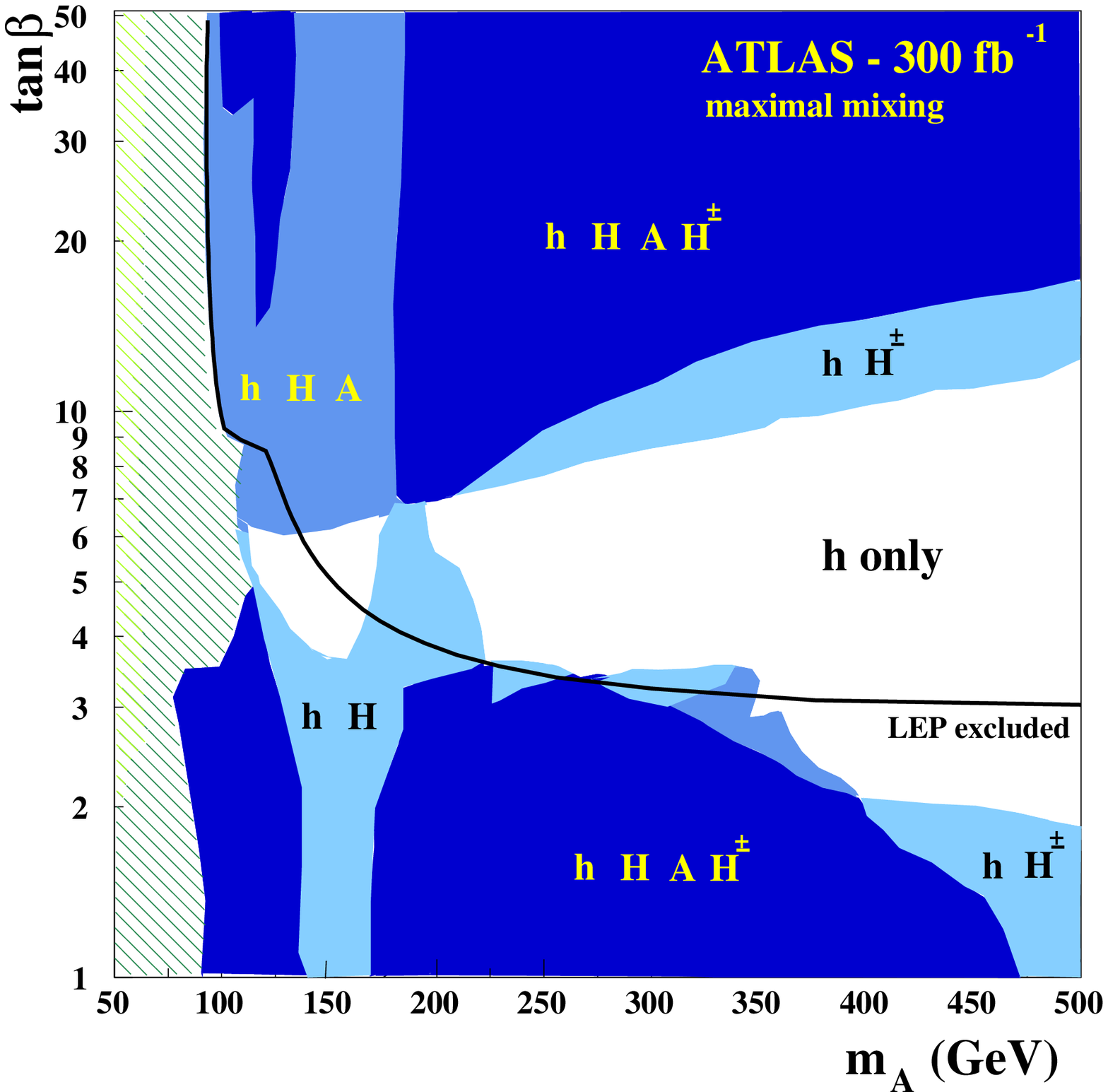}
%\centerline{
%            \psfig{figure=fig2.eps,width=6cm,height=6cm,angle=0}
%            \psfig{figure=fig4.ps,width=6cm,height=6cm,angle=0}}
\caption{The (non) unification of the three gauge couplings in the 
(SM) MSSM (left panel)  and the reach of LHC for the MSSM Higgs 
\protect\cite{tesla-tdr} (right panel) .\label{fig2}}
\end{figure}
This is illustrated in the left panel of 
Fig.~\ref{fig2}. It is imperative to find 'direct' 
evidence for SUSY. As  a result, SUSY searches have been an important benchmark
against which the capabilities and physics potential of the upcoming colliders
such as the LHC or the ones in future such as the ILC, have been evaluated.

The sparticle mass spectrum depends on the mechanism
responsible for SUSY breaking and can vary widely, but
the sparticle  spins and couplings are predicted  unambiguously.
With the help of the  the LHC and the ILC in the $e^+e^- mode$
~\cite{SUSYbook,tesla-tdr,atlas-tdr,lhc-ilc}
we hope to  find the sparticles, measure their masses spins and couplings.
The masses and the couplings of the $\tilde\chi^\pm, \tilde\chi^0_l$ and 
the supsesymmetric partners of the third generation of the quarks/leptons,
can depend on the SUSY breaking mechanism and parameters.  The LHC will be 
able to 'see' the strongly interacting sparticles if the SUSY breaking 
scale is TeV.  If the sparticle mass is within the kinematic reach of
the ILC, we should be able to make accurate mass measurements and  spin 
determination.  The LHC and the  ILC together can even help us determine the 
SUSY model parameters and hence the SUSY breaking mechanism~\cite{lhc-ilc}.
On this background it is important to inquire about the special role, if any,
that the PLC can play in this context. 

There are certain regions in the SUSY parameter space where the LHC and the 
ILC in the $e^+e^-$ mode  may be blind and or the signal may be lost.  The 
$\gamma \gamma$ mode and $e \gamma$ mode does provide possibilities to 
search for SUSY in this case. However, a more important question to ask is 
what are the unique possibilities offered by the PLC. Almost all of these come 
in the context of the Supersymmetric Higgs sector; especially in the context 
of Higgs sector with CP violation. PLC with its option of having
highly polarised photons, offers some unique possibilities. Some of these have
already been discussed in the meeting~\cite{peter,filip,piot}. In the next 
section we would discuss  these one by one.

\section{CP conserving MSSM Higgs sector and the PLC}
The PLC provides truly unique possibilities in probing the Higgs sector
in the MSSM~\cite{SUSYbook,abdel-hak}. In Supersymmetric theories there are 
(at least) five scalar states: $h,H,A$ and $H^\pm$.  $h,H$ are  CP even
whereas  $A$ is  CP odd and the $M_h$ is bounded from above.  In the decoupling 
limit $h$ will have properties very similar to a SM Higgs. The  MSSM  
parameters relevant for this sector are $\tan \beta$ 
(the ratio of vacuum expectation values), higgsino mass term $\mu$ and $M_A$.

The special features of a  $\gamma \gamma$ collider that are of special help,
 are:
\begin{enumerate}
\item Accurate measurements ($\sim 2\%$) of the $\Gamma_{\gamma \gamma}$ decay
width is possible.
\item
Polarisation of the laser and as well as that of the $e^+/e^-$ beam  can be 
tuned.
\item The $e \gamma$  option where polarised electron is 
scattered off  the high energy backscattered photon provides an extra channel.
\end{enumerate}

Below I will discuss three  examples where the PLC can  cover regions of SUSY 
parameter space which will be inaccessible to the LHC and the ILC in the 
$e^+e^-$ mode. 

\subsection{Higgs production through $\tau$--fusion mechanism}
Studies of the $\tilde\chi^+ \tilde \chi^-, 
\tilde\chi_j^0 \tilde \chi_i^0$ at $e^+e^-$ colliders provide possibilities 
of the  determination of SUSY parameters, $\mu, M_1, M_2$ and $\tan \beta$. 
However, accuracy of the $\tan \beta$ determination is degraded at 
large $\tan \beta$  mainly because the observable involves $\cos 2 \beta$.
A recent suggestion\cite{maggie1} is to use the $\tau$--fusion process
$\gamma \gamma \rightarrow \tau^+ \tau^- \phi \to \tau^+ \tau^- b \bar b$;
where $\phi$ denotes the Higgs boson.  
\begin{figure}
\includegraphics*[scale=0.30]{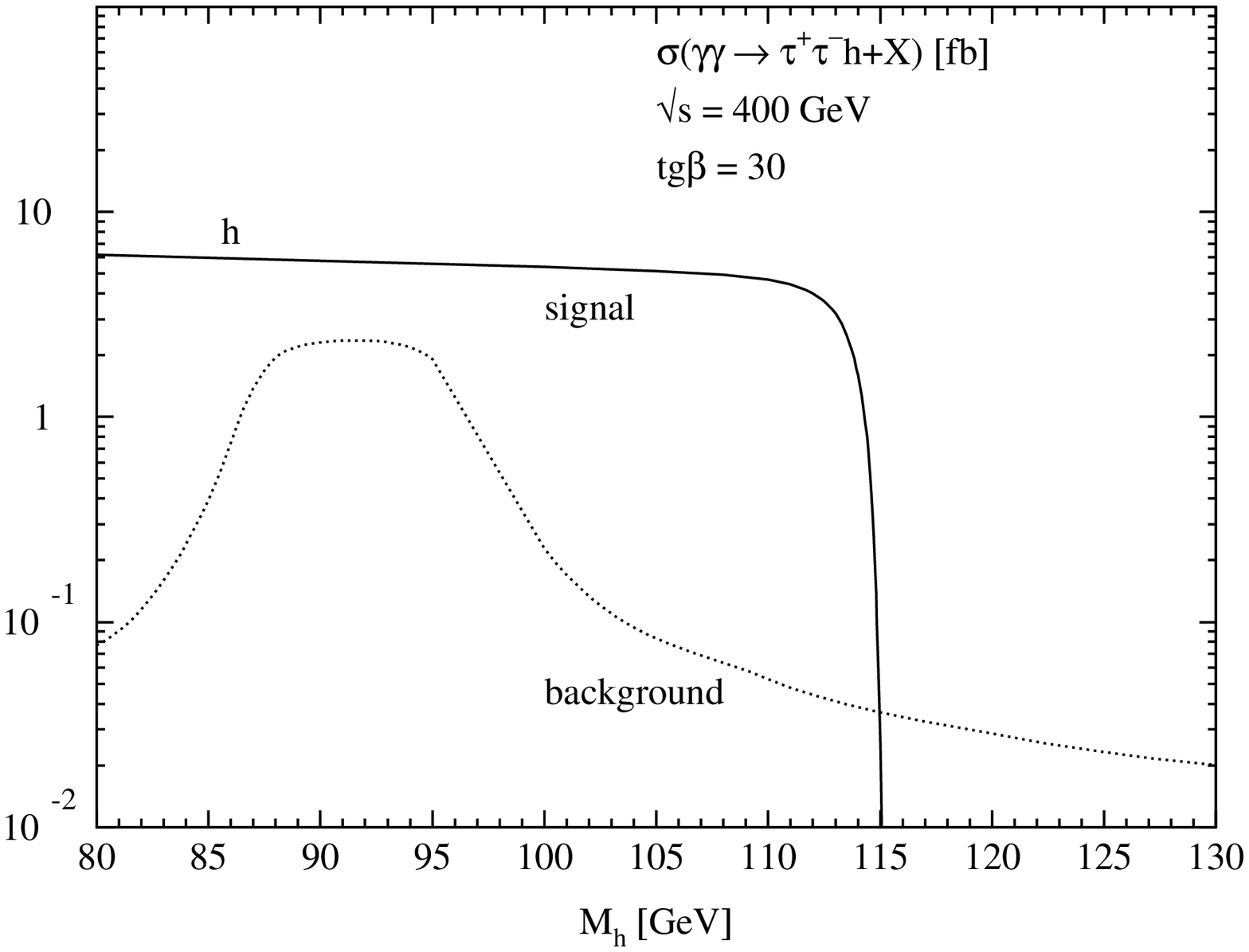}
\includegraphics*[scale=0.30]{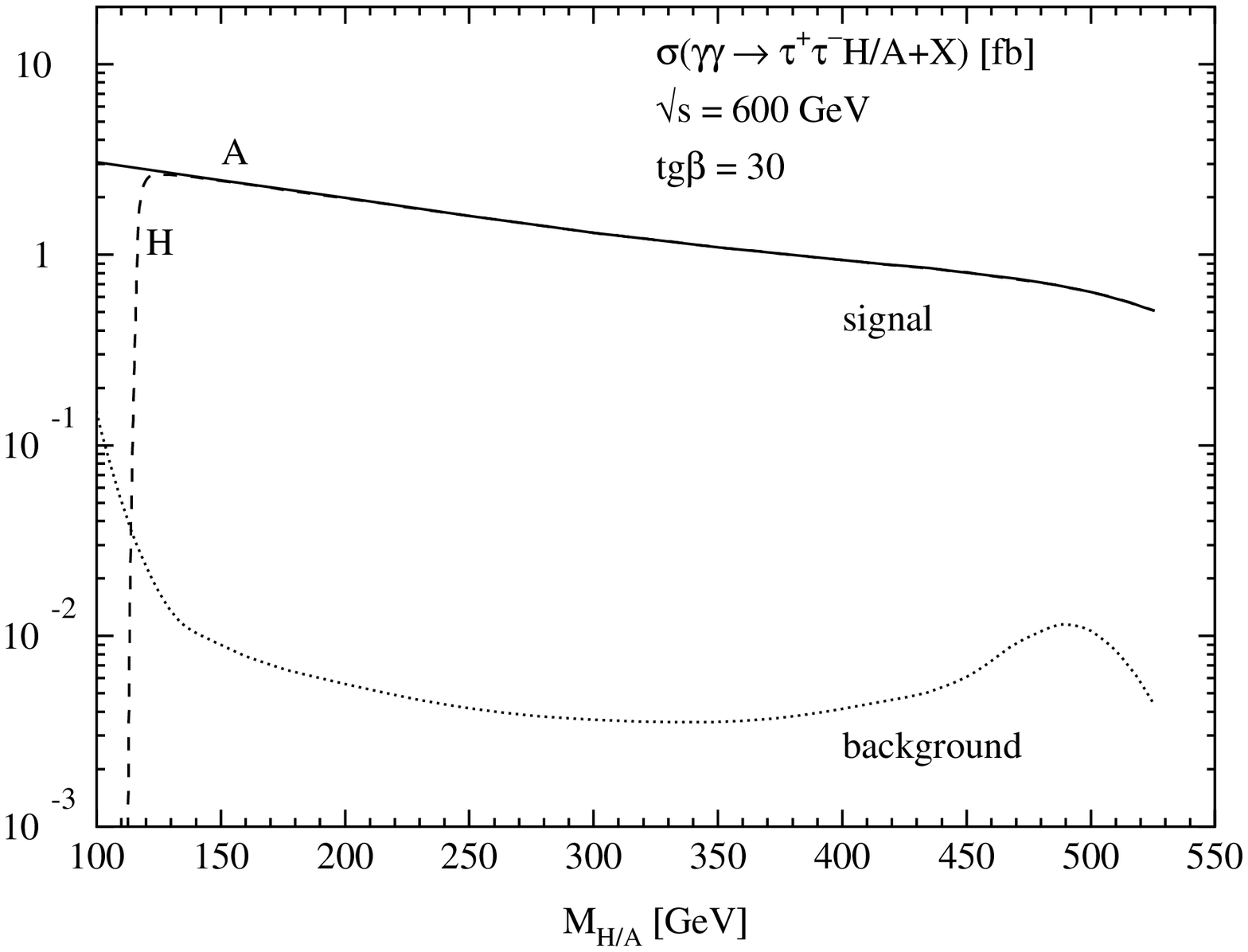}
\caption{$\tau$--fusion production rates for $h$ and $H/A$ production, 
along with the background  at the PLC 
shown in the left and the right panel respectively.  The peaked photon 
spectrum is used \protect\cite{maggie1}.\label{fig3}}
\end{figure}
Plots in Fig.~\ref{fig3} show that indeed for all the three Higgs states, the 
signal is substantially above the background. One can see that the process
offers a possibility of accurate $\tan \beta$ determination at large $\tan\beta$.
For example, at $\tan \beta = 30, \Delta \tan\beta = 0.9$--$1.3$. This  has
to be contrasted with the precision of $\Delta \tan \beta \sim 10$--$20$ 
that can be reached at the $e^+e^-$ option~\cite{ian}. The 
conclusions of this very interesting study need to be confirmed by simulations.

\subsection{Covering the LHC-Wedge for the MSSM}
As is seen in the right panel of Fig.~\ref{fig2}, 
in a plot taken from the TESLA-TDR~\cite{tesla-tdr},
for $\tan \beta \simeq 4-10$, $M_A, M_H > 200$--$250$ GeV, LHC will see only
one spin 0 state and the $H,A$ are not accessible for the first generation  
ILC.  
This region is referred to as the LHC-wedge. The $\gamma \gamma$ colliders 
offer unique possibilities of exploring this region. Since $H/A$ can be 
produced singly at a $\gamma \gamma$ collider, the reach in mass extends to 
$0.8 \sqrt{s}$ at the $\gamma \gamma$ option compared to the $0.5 \sqrt{s}$ 
at the $e^+e^-$ option. $\sqrt{s}$ of course is the cm energy of the parent 
$e^+e^-$ collider. The  QED background can be reduced by appropriately choosing
the laser photon and the electron helicities. For the larger $\tan\beta$ 
range, $b \bar b$ final state can be utilised effectively. However, for the 
smaller $\tan \beta$ values the $b
\bar b$ coupling reduces and the QED background being much higher for the
$t \bar t$ final state (due to the larger charge of the $t$ quark), it can 
not be used effectively either. In this region decays of $H/A$ into the
$\tilde \chi_1^\pm \tilde \chi_1^\mp, \tilde \chi_j^0 \tilde \chi_i^0$
may be used~\cite{maggie2}.

A detailed simulation of the $b \bar b$ final state for this LHC-wedge region
has been performed~\cite{maria2,maria1}. 
\begin{figure}
\psfig{figure=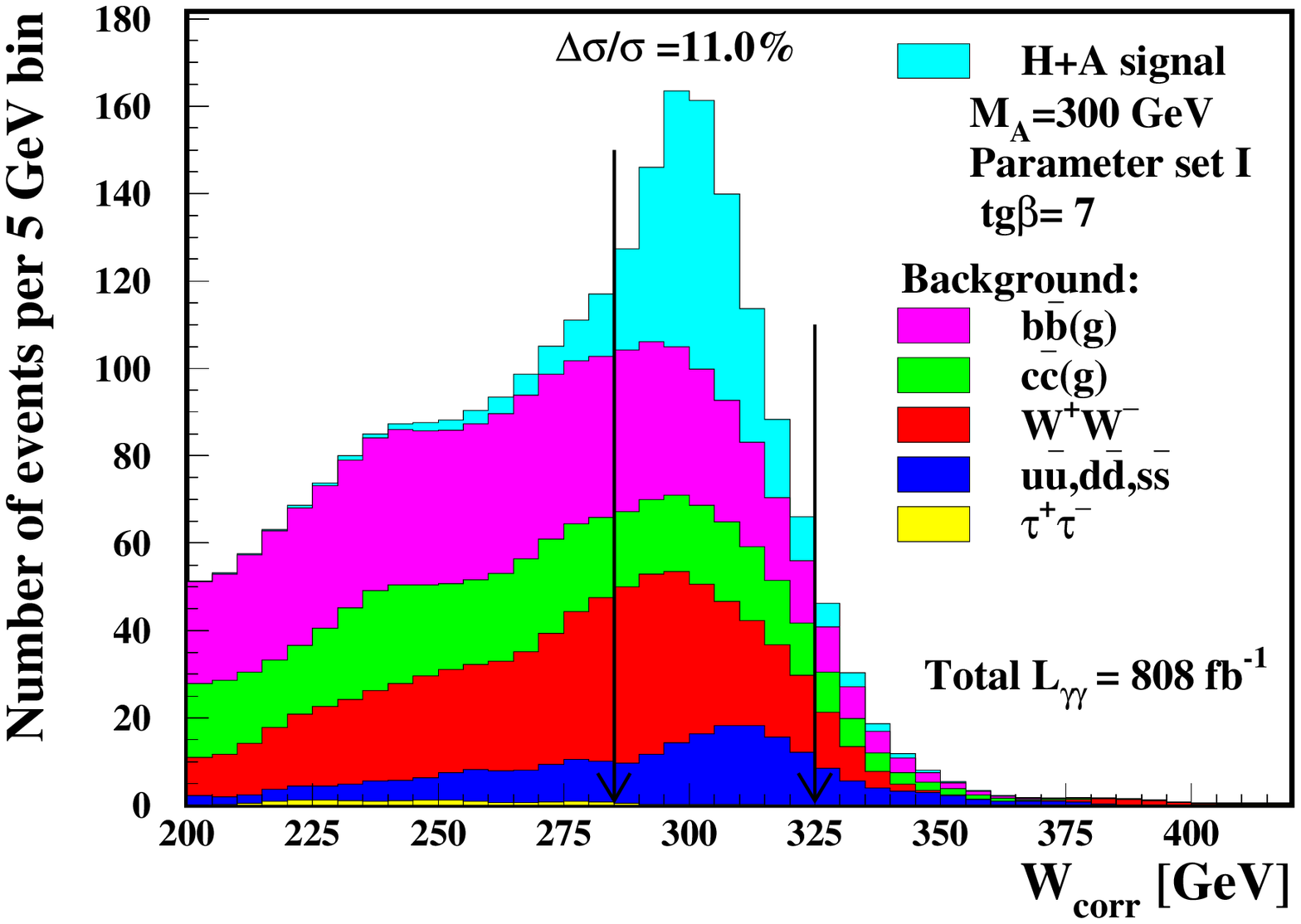,width=7cm,height=8cm,angle=0}
\psfig{figure=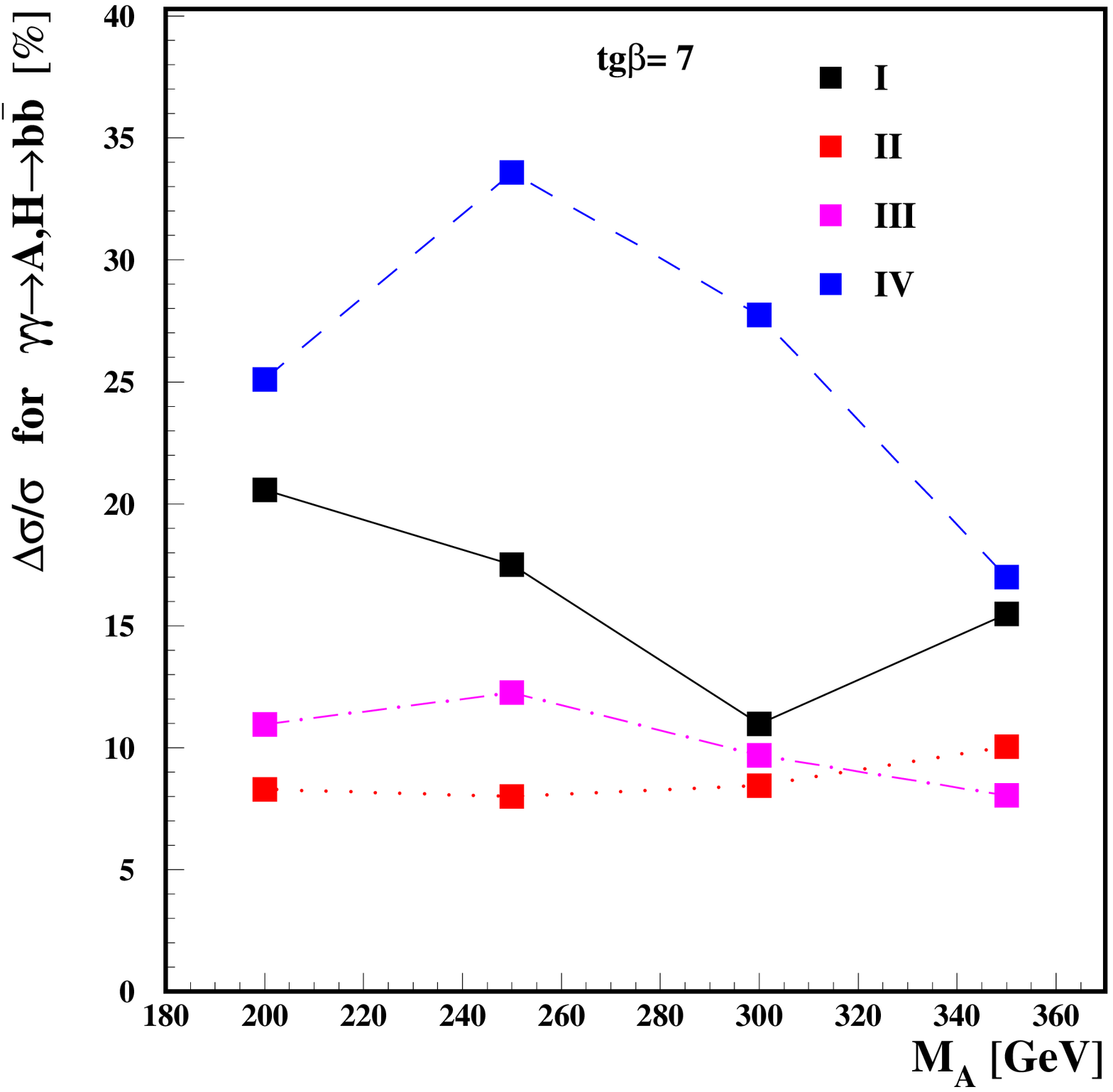,width=7cm,height=8cm,angle=0}
\caption{Precision possible in the measurement of $\gamma \gamma \rightarrow
H/A \rightarrow b \bar b$\protect\cite{maria1}.\label{fig5}}
\end{figure}
A summary of their conclusion is that for  the light Higgs the  $\gamma \gamma$ width can  be measured $\simeq 2 \%$, however in the case of $H/A$ the 
precision is somewhat worse : $\sim 11\%$--$21\%$. As said earlier, one can 
handle the QED background, by adjusting the helicities of the two photons.
For the $A/H$ there were suggestions to separate the two by choosing the  
polarisation vectors of the two photons to be  perpendicular and parallel.
However, in this case the QED background can not be handled easily. 

The precise measurement of the the width $\Gamma_{\gamma \gamma}$ at the PLC
can offer a probe of the contribution due to SUSY particles in the 
loop to the Higgs width~\cite{abdel-hak,abdelplb,usnpb}.

\section{CP-determination of the Higgs and the PLC}
CP violation in SUSY  used to be an embarrassment of riches, as there
exist large, number (44 to be precise) of phases of the SUSY parameters,
e.g. $\mu, A_f, M_i, i=1$--$3$ which can not be rotated away
by a simple redefinition of the fields. These can generate  unacceptably 
large electric dipole moments for fermions and hence one of
the solutions normally was to fine tune all the $\cp$ phases in SUSY to 
zero.  It has been shown that it is  possible for some combination of 
these phases to be {\cal O} (1) and yet satisfy {\it all} the constraints on 
the EDM's provided the first two generation of squarks are heavy~\cite{cpmssm}.
It has been demonstrated that such $\cp$ phases can induce CP mixing in the
Higgs sector of the MSSM~\cite{pilaftsis1,choi1,wagner1}. This leads to mixing 
between the CP-even $h,H$ and the CP-odd $A$ in the MSSM. The couplings 
of the three mass eigenstates $\phi_1,\phi_2,\phi_3$, ($m_{\phi_1} < m_{\phi_2} 
<m_{\phi_3}$), are modified compared to the CP-conserving case. In particular,
the $\phi_1$ may develop a large pseudoscalar component, giving 
$g^{VV\phi_1} < g^{VVH_{SM}}$ and hence $\sigma(e^+e^-\to Z^* \to 
Z \phi_1) < \sigma(e^+e^-\to Z^* \to Z h_{SM})$. This is the simplest way in
which CP-violation may invalidate the lower limits on the Higgs mass obtained
by LEP.  The LEP data can now allow a much lighter Higgs with a mass
$\lts 40$--$50$ GeV \cite{carena,Abbiendi:2004ww,susylim} due to a
reduction in the $\phi_1 ZZ$ coupling in the CPX scenario~\cite{wagner1}. The
latter 
corresponds to a certain choice of the CP-violating SUSY parameters, chosen so
as to showcase the CP-violation in the Higgs sector in this case.
In a large portion of this region all the usual search channels of such a light
Higgs at the LHC are also not expected to be viable~\cite{carena} due to the
simultaneous reduction in the coupling of the Higgs to a vector boson pair as
well as the $t\bar t $ pair. As a matter of fact for $\tan \beta : 3-5,
M_{H^+} : 50-100$ GeV,  there may exist a hole in the SUSY parameter space 
in case of CP-violation. Part of this hole can  be filled up, by  taking
advantage of the light $H^+$ which can be produced in the top decay and 
which in turn has a large branching ratio in the $\phi_1 W$ 
channel~\cite{dilip}. Even after this, some part of this 'hole' still remains.

A PLC will  be  able to produce such a neutral Higgs in all cases; 
independent of whether it is a state with  even/odd or indeterminate CP parity.
It is possible to determine the CP mixing, if present, by using the 
polarisation of the initial state $\gamma$ or that of the fermions into 
which the  $\phi_i$ 
decays~\cite{Grzadkowski:1992sa,Asakawa:1999gz,asakawa1,ritesh1,ellis1,zerwas1}.
A unique feature of a PLC is that the two photons can form a $J_z = 0$ state
with both even and odd CP. As a result a PLC has a similar level of
sensitivity for both the CP-odd and CP-even components of a CP-mixed state:
\begin{equation}
  {\rm CP\!-\!even:}
  \epsilon_1\cdot \epsilon_2 = -(1+\lambda_1\lambda_2)/2 , \quad
  {\rm CP\!-\!odd:}
  [\epsilon_1 \times \epsilon_2] \cdot k_{\gamma}
  =\omega_{\gamma} i \lambda_1(1+\lambda_1\lambda_2)/2,
\end{equation}
$\omega_i$ and $\lambda_i$ denoting the energies and  helicities of the
two photons respectively; the helicity of the system is equal to
$\lambda_1-\lambda_2$.
This contrasts the $e^+e^-$ case, where it is easy to discriminate between
CP-even and CP-odd particles but may be difficult to detect small CP-violation
effects for a dominantly CP-even Higgs boson~\cite{Hagiwara:2000bt,lhc-ilc,CP-report}.
For the PLC, one can form three polarization asymmetries in terms of helicity
amplitudes which give a clear measure of CP mixing \cite{Grzadkowski:1992sa}.
Note however that these require linearly polarised photons in addition to the 
circularly polarised photons.
One can also use information on the decay products of $WW$, $ZZ$,
$t\bar t$ or $b\bar b$ coming from the Higgs decay.
Even with just the circular beam polarization almost mass degenerate
(CP-odd) $A$ and (CP-even) $H$ of the MSSM may be separated
\cite{maggie2, maria2,maria1,Asakawa:1999gz}.  In  the situation that the mass
difference between the H and A is less than the sum of their widths, a coupled 
channel analysis technique~\cite{aposto2} has to be used. The authors of Refs.
~\cite{ellis1} and \cite{zerwas1} explore this situation whereas the use of 
decay fermion polarisation for determination of the Higgs CP property for a
generic choice of the MSSM parameters is explored in Ref.~\cite{ritesh2}.

The process $\gamma \gamma \rightarrow f \bar f$ receives contribution from the
process where the $\phi$ is exchanged in the $s$--channel and thus probes the
$\phi \gamma \gamma$ and $\phi f \bar f$ couplings:
$$
{\cal V}_{f \bar f\phi} = -ie\frac{m_f}{M_W} \left(S_f+
i\gamma^5P_f\right)$$
 and 
$$
{\cal V}_{\gamma\gamma\phi} = \frac{-i\sqrt{s}\alpha}{4\pi}\left[S_{\gamma}(s)
\left(\epsilon_1.\epsilon_2-\frac{2}{s}(\epsilon_1.k_2)(\epsilon_2.k_1) \right)
\right.
-\left. P_{\gamma}(s)\frac{2}{s}\epsilon_{\mu\nu\alpha\beta}\epsilon_1^{\mu}
\epsilon_2^{\nu} k_1^{\alpha} k_2^{\beta} \right].$$
$\{S_f,P_f,S_{\gamma},
P_{\gamma}\}$ depend upon $m_{H^+}, \ \tan\beta, \ \mu, \ A_{t,b,\tau}$, 
$\Phi_{t,b,\tau}, \ M_{\tilde q}, \ M_{\tilde l}$ etc. in ($CP$ violating) MSSM.
The helicity amplitudes involve four CP-even and CP-odd combinations of the
different form factors, $x_i,y_i,i=1,4$, respectively. The 
QED background is  $P$, $CP$ and chirality conserving, while the 
$\phi$  exchange diagram violates these symmetries. Thus   
nonzero values of $\{x_i,y_j\}$ indicate existence of chirality flipping 
interactions as opposed to the chirality conserving QED interactions.
As a result the fermion polarisation can be a probe of the $\phi$
contribution as well as any possible CP-violation in the $\phi \gamma \gamma$ 
and $\phi t \bar t$ coupling. The polarisation of the initial state $\gamma$
can be controlled by adjusting the initial laser and the $e$ polarisation. The
$\phi$ contribution is enhanced using the combination $\lambda_e \times 
\lambda_l = -1$. One can construct observables, with unpolarised and polarised
laser and $e$ beams: $P_f^{U}$ and $\delta P_f^{CP} = P^{++}_f + P^{--}_f$ 
which are both probes  of CP violating interaction, and 
$\delta P_f^+ = P^{++}_f - (P^{++}_f)^{QED}$ $\delta P_f^- = 
P^{--}_f - (P^{--}_f)^{QED}$ which are probes of chirality flipping 
interactions. Here $+/-$ refer in the (double) superscripts of $P_f$
refer to the polarisation of the $e, \lambda_e$. 
\begin{figure}[htb]
\centerline{
            \psfig{figure=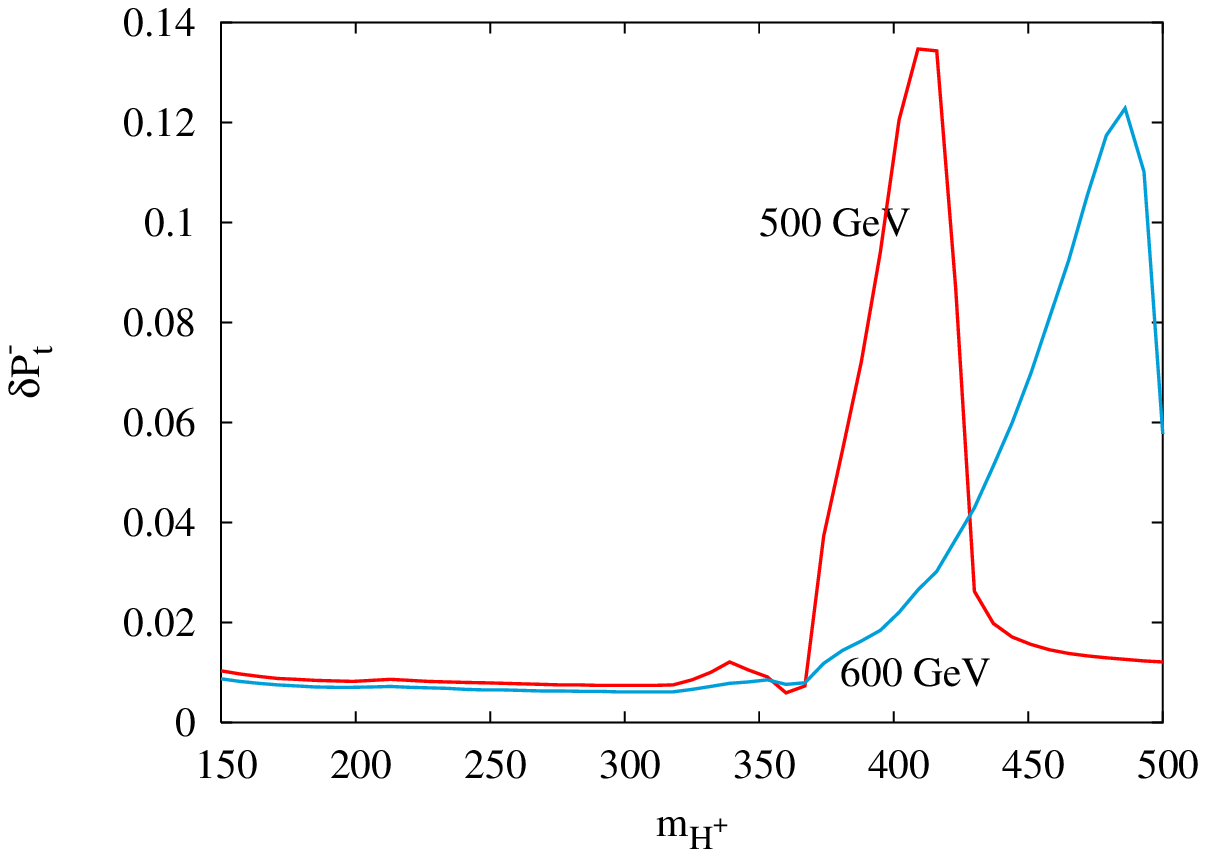,width=7cm,height=7cm,angle=0}
            \psfig{figure=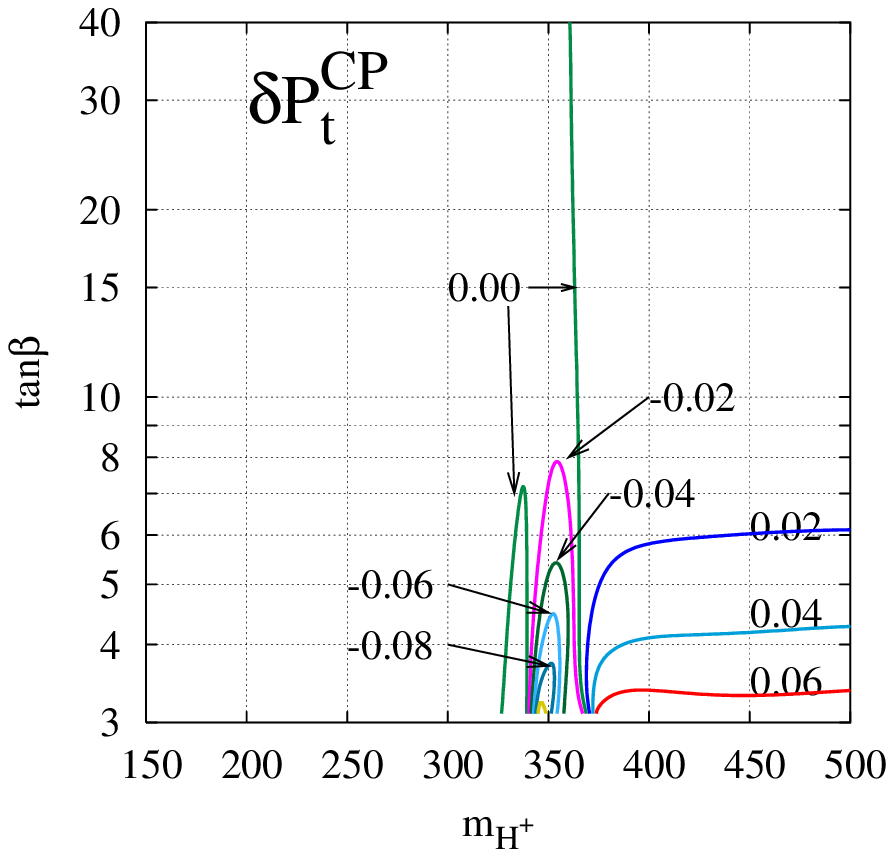,width=7cm,height=7cm,angle=0}
                   }
\caption{Left panel shows $\delta P_t^-$ 
as a function of $m_{H^+}$ for $E_{cm}=$ 500 GeV and 600 GeV, while the right 
panel shows $\delta P_t^{CP}$ over the $\tan \beta$--$m_{H^+}$ plane for 
CP violating phase $\Phi = 90^\circ$, in the CPX scenario~\cite{wagner1}.
\label{fig6}}
\end{figure}
Left panel of Fig.~\ref{fig6} shows the predicted value of $\delta P_t^-$ as a 
a function of $m_{H^+}$ for $E_{cm}=$ 500 GeV and 600 GeV, using the 
ideal back-scattered photons, with  $x_c =4.8$. The  peak occurs when
the average  mass of $\phi_2$ and $\phi_3$ ($\bar m_{\phi}$ matches with the 
$\sqrt{s}_{\gamma \gamma}$ value where the backscattered laser photon 
luminosity peaks. The right panel shows, the expected values of the CP-violating
asymmetry $\delta P_t^{CP}$ as a function of the two MSSM parameters, 
$\tan \beta$ and $\mu$, for $\Phi = 90^\circ$. $E_{cm}$ is adjusted for
each point in the  scan such that the peak of the photon spectrum matches with
scaled mass $\bar m_{\phi}$. Nowhere in this range of the parameters are the
two states extremely degenerate, and hence a coupled channel analysis is
not required.  We see that even in this case, the size of the expected 
asymmetries is not too small. Thus in a generic case of CPV MSSM the PLC
can probe this CP-mixing in the Higgs sector.

The case of extreme degeneracy has been studied for the PLC 
in~\cite{ellis1,zerwas1}
\begin{figure}[htb]
\includegraphics*[scale=0.5]{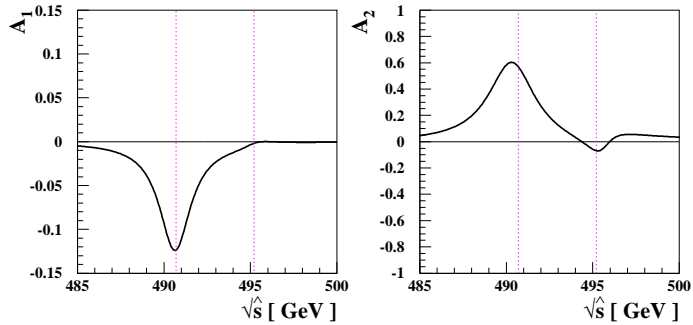}
\caption{The asymmetries with respect to photon helicities\protect\cite{ellis1}.
  \label{fig7}}
\end{figure}
Fig.~\ref{fig7} shows the CP violating asymmetries:
$$
{\cal A}_1\equiv
\frac{\hat\Delta_1}{\hat{\sigma}_{++}+\hat{\sigma}_{--}}\,,
{\cal A}_2\equiv
\frac{\hat\Delta_2}{\hat{\sigma}_{+-}+\hat{\sigma}_{-+}}.
$$
Here $\hat{\Delta}_i, i=1,2$ have been constructed out of cross-sections
with final quarks in different helicity states. $+/-$ refer to photon and 
the final state quark  helicities. For the
chosen values of the parameters, the asymmetries are sizable only near the
$\phi_2$ mass.

The analysis of Ref.~\cite{zerwas1} investigates the asymmetries constructed
using linearly polarised photons. 
\begin{figure}[htb]
\includegraphics*[scale=0.5]{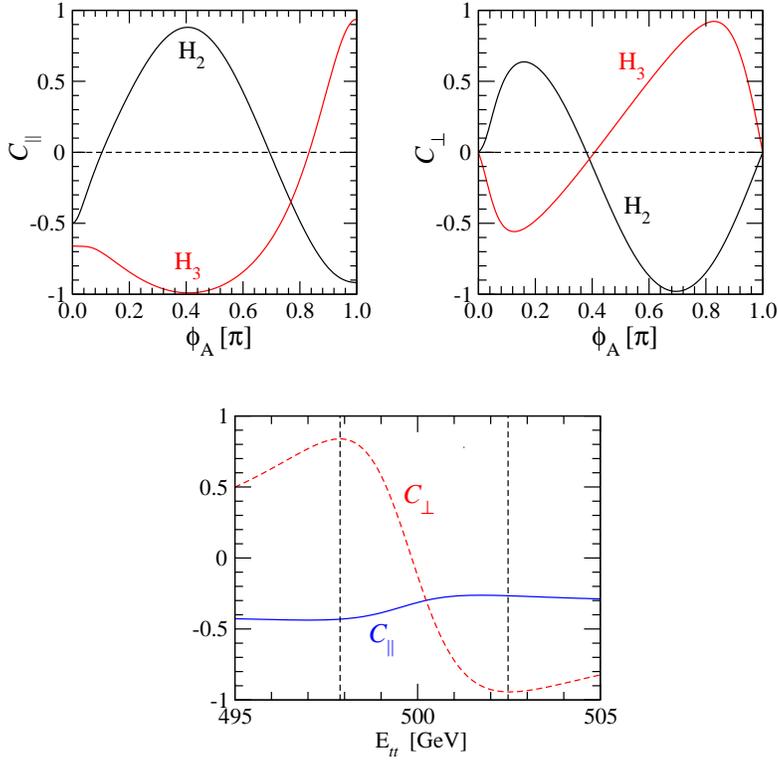}
\caption{CP violating correlators constructed using linearly polarised
photons, as a function of the CP violating phase $\Phi_A$ and the $t \bar t$
centre of mass energy.\protect\cite{zerwas1}. 
\label{fig8}}
\end{figure}
Fig.~\ref{fig8} taken from Ref.~\cite{zerwas1} shows the correlators 
\begin{eqnarray}
&&{\cal C}_\parallel = -\, \frac
   {2\,\real \sum \langle+,\lambda\rangle
                  \langle-,\lambda\rangle^*}
   {\sum \left(|\langle+,\lambda\rangle|^2
                      +|\langle-,\lambda\rangle|^2\right)}
   \\[2mm]
&&{\cal C}_\perp     =+\, \frac
   {2\, \imag\sum \langle+,\lambda\rangle\langle-,\lambda\rangle^*}
   {\sum \left(|\langle+,\lambda\rangle|^2
                      +|\langle-,\lambda\rangle|^2\right)}
\end{eqnarray}
as a function of the CP-violating phase $\Phi_A$ and the $t \bar t$ centre of 
mass energy.

The decay leptons from $t$-quark carry information about its polarization. 
One can construct asymmetries combining charge of lepton and polarization of 
the initial state $e^-$ of the PLC.
Parameterizing the cross-sections  as $\sigma(\lambda_{e^-}, Q_\ell)$, one can
define mixed charge-polarisation asymmetries:
$$ {\cal A}_1 = \frac{\sigma(++)-\sigma(--)}{\sigma(++)+\sigma(--)}\hspace{1cm}
   {\cal A}_2 = \frac{\sigma(+-)-\sigma(-+)}{\sigma(+-)+\sigma(-+)}$$
$$ {\cal A}_3 = \frac{\sigma(++)-\sigma(-+)}{\sigma(++)+\sigma(-+)}\hspace{1cm}
   {\cal A}_4 = \frac{\sigma(+-)-\sigma(--)}{\sigma(+-)+\sigma(--)}$$
\begin{figure}[htb]
\includegraphics*[scale=0.8]{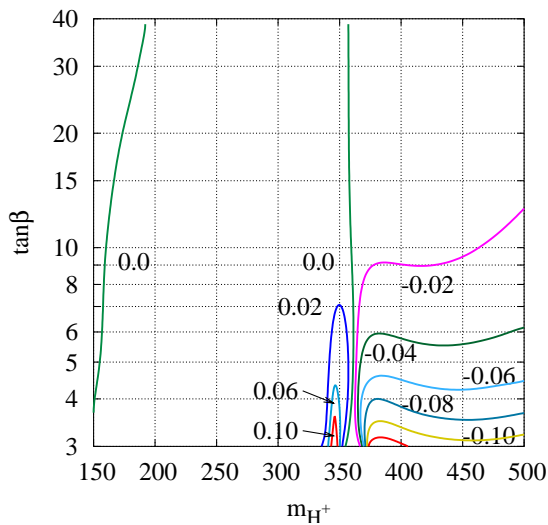}
\caption{Asymmetries using the initial state $\gamma$ polarisation and final 
state lepton charge~\protect\cite{ritesh2}.
\label{fig9}}
\end{figure}
Fig~\ref{fig9} shows these asymmetries over the  $\tan \beta$--$m_{H^+}$
plane, for CPX scenario, for $\Phi = 90^\circ$, for a fixed beam energy. 
We see that the size of asymmetries goes as high as $10 \%$ and  tracks the
polarisation asymmetries shown in the earlier figures. 

\section{$e \gamma$, $\gamma \gamma$  collider and sparticle production.}
Production of the sparticles, at the $\gamma \gamma$ 
option~\cite{Goto:1990ub,Cuypers:1992tj,Koike:1995be,Ghosh:2001iu} 
as well as that at the $e$-$\gamma$
option~\cite{Kon:1992vg,Goto:1992kr,Freitas:2005et} has been studied.
The interesting thing about charged sparticle production at the 
$\gamma \gamma$ colliders is that the cross-sections, to leading order,  
are entirely determined by their charge and mass, as compared to the case of 
an $e^+e^-$ collider where the cross-section may  depend on the 
various mixing angles due to the presence of the  weak gauge bosons.
 This property could provide us complementary information about the
models, e.g., universality of the masses for sleptons and squarks in the
1st and 2nd generations. It should be emphasized that, as the 
$\gamma\gamma$ cross sections involve an $s$-wave contribution, 
they will be much larger than that of $e^{+}e^{-}$
if $\sqrt{s}$ is large compared to the mass threshold.  In the $e$--$\gamma$
option, a charged sfermion can  be produced in association
with a neutral gaugino or vice versa.  If the mass difference between the two
is large, then this offers a higher kinematical reach compared to the $e^+e^-$ 
option. Further, use of polarisation allows to enhance the signal. Again,
in this case the dependence of the cross-section on the SUSY parameters is 
reduced. For example,  even in the case of (say) $\tilde \chi_1^0$ produced 
in association with a $\tilde e_R$, the production will involve only the Bino 
component of the $\tilde \chi_1^0$.  The threshold dependence of the 
$\tilde \nu \tilde\chi_1$ production may be used 
for the determination of the sum of the two masses and hence can afford a good
determination of the $\tilde \nu$ mass~\cite{Freitas:2005et}. Single
sneutrino production can be used to study SUSY at the PLC in the R-parity 
violating scenarios quite effectively~\cite{Ghosh:2001iu}.

\section{Conclusion}
Thus we see that a PLC can offer a chance of real improvements in the
accuracy $\Delta \beta$ of $\tan \beta$ measurements at large $\tan \beta$ 
using $\tau \tau$ fusion. A PLC also provides major gains for the SUSY 
Higgs sector as it gives a reach for $H/A$ in regions where the LHC does 
not have any. Further, the $s$ channel production mode increases the reach 
by a factor $\sim 1.6$ compared to the $e^+e^-$ option.  The advantages of a 
$\gamma \gamma$ collider are even more if CP violation is present.
Polarisation asymmetries constructed using initial
state photon polarisation and final state fermion polarisations, can be a
very good probe of the CP violation in the Higgs sector. LHC/ILC are not
very capable when it comes to  probing CP-mixing.  
The $H/A$ contribution can be probed through mixed
polarisation-charge asymmetries, i.e asymmetries in initial state polarisation
and final state lepton charge.
If CP violation makes the lightest Higgs
dominantly pseudoscalar and hence 'invisible' at LEP/ILC/LHC then
$\gamma \gamma$ collider is the only place it can be produced
For SUSY searches, $\gamma \gamma$  and $e \gamma$ collider
can offer some interesting possibilities  for sneutrino and gaugino
searches; particularly the production cross-sections are independent of the
the different mixing angles for the charged sparticles. Further the high 
polarisation of the  backscattered laser can be put to very good use.

\section{Acknowledgment}
It is a pleasure to thank the Organisers for the wonderful organisation
and the atmosphere at the meeting.

\end{document}